Vincenzo Caligiuri, Hyunah Kwon, Andrea Griesi, Yurii P. Ivanov, Andrea Schirato, Alessandro Alabastri, Massimo Cuscunà, Gianluca Balestra, Antonio De Luca, Tlek Tapani, Haifeng Lin, Nicolò Maccaferri, Roman Krahne, Giorgio Divitini, Peer Fischer and Denis Garoli *

# Dry synthesis of bi-layer nanoporous metal films as plasmonic metamaterial

**Abstract:** Nanoporous metals are a class of nanostructured materials finding extensive applications in multiple fields thanks to their unique properties attributed to their high surface area and interconnected nanoscale ligaments. They can be prepared following different strategies, but the deposition of an arbitrary pure porous metal is still challenging. Recently, a dry synthesis of nanoporous films based on the plasma treatment of metal thin layers deposited by physical vapour deposition has been demonstrated, as a general route to form pure nanoporous films from a large set of metals. An interesting aspect related to this approach is the possibility to apply the same methodology to deposit the porous films as a multilayer. In this way, it is possible to explore the properties of different porous metals in close contact. As demonstrated in this paper, interesting plasmonic properties emerge in a nanoporous Au-Ag bi-layer. The versatility of the method coupled with the possibility to include many different metals, provides an opportunity to tailor their optical resonances and to exploit the chemical and mechanical properties of components, which is of great interest to applications ranging from sensing, to photochemistry and photocatalysis.

## 1 Introduction

Nanoporous metals (NPMs), are a unique class of metamaterials, made of solid metals with nanosized porosity, ultrahigh specific surface area, good electrical conductivity, high structural stability, and tunable optical properties.[1] These unique characteristics have facilitated their use in many fields, such as electrochemical and optical sensing, photo and chemical catalysis, and advanced energy technology. In particular, NPMs have been extensively explored for plasmonic applications,[2] as their structural and optical properties can be easily tuned by controlling the conditions used during their preparation. Many different methods, such as templating, dealloying, and colloidal chemistry, have previously been employed to prepare NPMs, each with its own merits and limitations. Template-based methods enable a precise control over size and microstructure, but are generally difficult and time consuming to implement.[3] Wet-chemistry methods have been also explored, however, the control over the porosity is still a challenge. Dealloying, a widely used method, allows varied pore sizes and structural properties, but it faces challenges in completely removing less noble metals during the etching process, impacting the final NPM's properties. More importantly, it works only for a limited number of metals. [4]–[9] To address these limitations, a new dry-synthesis route to obtain impurity-free NPM films has been recently demonstrated.[10] In this method, the NPMs are formed by the coalescence of metal nanoparticles in a low-temperature plasma, enabling the production of a large set of metals with no impurities. The films can be extremely thin (about 10 nm), and can be thickened through replicating the process multiple times, i.e. thicker films can be deposited (as *n* layers) replicating the process several times using the *n-1* layer as substrate. More importantly, the process allows for the serial deposition of multi-layered materials, either of the same or different metals. Therefore, it is particularly advantageous for creating bi-layer (or multi-layer) NPMs that are in close contact, which is challenging with conventional techniques. Here, we focus on bi-layer NPMs using gold and silver, the most used plasmonic materials, and we analyze their optical properties in detail. Our system is one of the first examples of bimetallic nanoporous coating, and promises a number of potential applications in various fields.

The use of bi-metallic nanostructures has been recently reported as a very interesting configuration to explore plasmonics, in particular, for catalysis. Compound nanoparticles (NPs) are typical structures, where a plasmonic metal (generally Au or Ag) is supported by a catalytic metal (Pd, Pt, etc.) to leverage the combined effect of plasmonics that converts optical radiation into energy, and traditional catalysis that promotes adsorption of molecules and that facilitates chemical conversion due to surface active atoms. Bimetallic NPs systems have been explored in several recent papers [11][12]. Among them, the so-called "antenna-reactor" systems constitute the simplest but most effective example of bi-metallic nano-structures for catalysis.[12][13][14] [15] Bi-metallic nano-structures find also broad applications in plasmonic heat generation,[17] Surface Enhanced Raman Spectroscopy (SERS),[18]–[20] dechlorination,[21] photothermal therapy,[22] among the others. Either a bottom up (chemical synthesis) or a top-down (physical or chemical vapor deposition, lithography, etc.) approach is adopted. All these techniques, involve sophisticated procedures, relying strongly on spatial order. The approach described in this work relies instead on a macroscopic homogenization effect for the achievement of a tailored dielectric permittivity. Our approach, therefore, lifts the strict geometric constraints of previous methods, and thereby expands

*Corresponding author: Denis Garoli, Dip. di Scienze e Metodi dell'Ingegneria, Università di Modena e Reggio Emilia, via Amendola 2, 42122 Reggio Emilia, Italy , E-mail: denis.garoli@unimore.it

**Vincenzo Caligiuri, Denis Garoli, Roman Krahne, Andrea Griesi, Yurii P. Ivanov, Giorgio Divitini**: Istituto Italiano di Tecnologia, Via Morego 30, 16136 Genova, Italy

**Hyunah Kwon, Peer Fischer**: Institute for Molecular Systems Engineering and Advanced Materials, Heidelberg University, 69120 Heidelberg, Germany; Max Planck Institute for Medical Research, 69120 Heidelberg, Germany

**Andrea Schirato**: Deparment of Physics, Politecnico di Milano, Piazza L. da Vinci 32, I-20133 Milan, Italy

**Alessandro Alabastri**: Department of Electrical and Computer Engineering, Rice University, 6100 Main Street MS-378, Houston, TX 77005, USA

**Massimo Cuscunà, Gianluca Balestra**: Institute of Nanotechnology - CNR NANOTEC  c/o Campus Ecotekne, Via Monteroni  73100 Lecce (Italy)

**Tlek Tapani, Haifeng Lin, Nicolò Maccaferri**: Department of Physics, Umeå University, Linnaeus v€ag 24, SE-90187 Umeå*, Sweden

**Vincenzo Caligiuri, Antonio De Luca**: 1. Dipartimento di Fisica, Università della Calabria, via P. Bucci 33b, 87036 Rende (CS), Italy.

2. Istituto di Nanotecnologia (CNR-Nanotec) SS di Rende, via P. Bucci 33c, 87036 Rende, Italy.

the application space of possible bi-metallic random porous nano-structures.

## 2 Results and discussion

The dry-synthesis that has been recently demonstrated, [10] is based on the plasma treatment of a dense layer of nanoparticles deposited on a sacrificial thin PMMA polymer film. The metal layer is deposited by e-beam evaporation at an oblique angle between 70° and 80° on top of the sacrificial layer, which is then removed by means of plasma etching. While stable metals like Au can be obtained using oxygen ($O_2$) plasma, other metals prone to oxidation such as Ag need to be prepared by etching the sacrificial PMMA layer using a not-reactive plasma (like Ar or $N_2$). We verified that the procedure works nicely for a wide range of metals including Au, Ag, Cu, Pt, Pd, Al, and Ni. The obtained porous morphology can be significantly different among the different metals. A single run could not create more than 14 nm of metal, while it is easy to deposit thick films in $n$ successive processes where the sacrificial PMMA layer is re-deposited on top of the previously deposited porous ($n-1$) layer and the process replicated. By applying this multi-layer approach, it is also possible to deposit successive layers of different metals. In particular, as illustrated in Fig. 1, we fabricated a bi-layer of nanoporous Au with a top layer of nanoporous Ag. The obtained morphology can be

Figure 2a reports a schematic of the bi-layer geometry considered in the model, with Fig. 2b showing a top view of the film. By solving Maxwell's equations locally, our numerical simulations provide an estimation of the electromagnetic field distribution along the surfaces of the two (the top and bottom) porous layers. Figures 2c and 2d depict some exemplary results for a wavelength of 400 nm impinging on the hetero-metallic bi-layer (Ag on top of Au), exhibiting strong field localisations at the deep subwavelength scale and intense field enhancement factors at the metal discontinuities and nano-asperities across the porous surfaces. Interestingly, the use of two distinct materials in the bilayer structure introduces differences in the overall electromagnetic and optical response of the porous system compared to the homo-metallic counterpart. To illustrate this effect, we repeated our simulations on the same bi-layer structure, yet by using as example the Ag optical properties for both the top and the bottom layers. The results of our complementary simulations are summarized in Fig. 2e. In comparing the values of optical absorptivity computed for the two cases (experimental hetero-metallic Ag-Au and a fictive homo-metallic Ag-Ag system), major discrepancies are predicted. In particular, the experimental Ag-Au layer allows for a larger contrast in absorptivity both spatially (between the top-Ag and bottom-Au layers) and in wavelength (e.g., between 400 nm and 550 nm). At λ=400 nm, while the Ag-Ag case displays comparable absorptivity in both layers (between 15 directly compared with the morphologies of the individual metals (Fig. 1A, Au; Fig. 1B, Ag). As can be observed, the bi-layer looks like a sandwich structure of the two NPMs, therefore producing a bimetal porous junction.

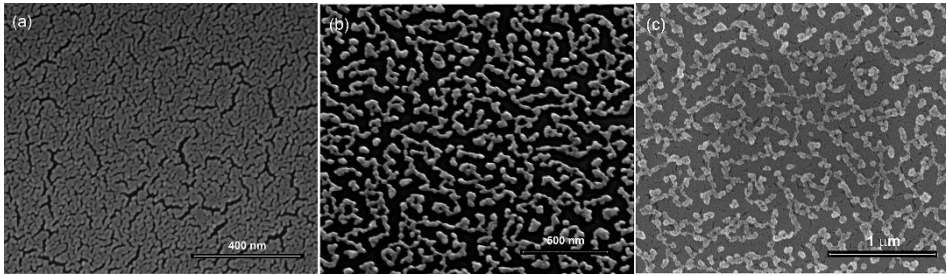

**Fig. 1:** SEM micrographs of the prepared porous films. (a) Nanoporous Au; (b) Nanoporous Ag; (c) bilayer Au-Ag.

A numerical study of the electromagnetic response of the porous hetero-metallic system can be an important complement to the experimental analyses. For an accurate description of the nanometric features of the layers, the SEM images of the sample were directly employed in our Finite Element Method (FEM)-based model (developed using the commercial software COMSOL Multiphysics). Specifically, top views of the metal films were imported into our solver and used to build the numerical geometry used in the simulations via extrusion along the thickness of the layer (20 nm). Our approach, detailed in previous works[7], [8], [23], enabled us to numerically account for the actual discontinuities and profile of the materials down to the nanoscale (by approximating the same porosity in the direction of the film thickness). A linearly-polarised monochromatic light at normal incidence was considered, and an infinite layer in the plane was assumed for numerical reasons (with no significant impact on the overall electromagnetic behaviour of the system, provided a sufficiently large porous unit cell is considered. Au and Ag's optical properties were taken from [24], respectively, for the two layers.

% and 20 %) the Ag-Au case features an absorptivity in the top layer about ten times larger than in the bottom one. At λ=550 nm, instead, the spatial distribution is partially reversed for the hetero-metallic case, with the bottom layer dissipating almost twice the top layer. The homo-metallic case shows again similar dissipation values. These differences between hetero- and homo-metallic systems can be ascribed to the electric permittivity values affecting the coupling between layers. This effect is amplified at shorter wavelengths where the Au interband transition plays a more relevant role. Overall, such peculiar electromagnetic responses highlight the potential of utilizing different nanoporous materials stacked in thin coupled layers to achieve engineered dissipation patterns.

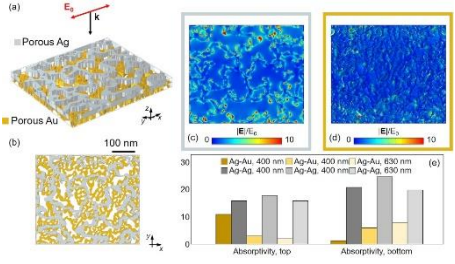

**Fig. 2:** (a) Numerical hetero-metallic bi-layer NPM used in the simulation. (b) top view of the bi-layer NPM. (c-d). Computed electromagnetic field enhancement evaluated at the upper surface of the top (Ag, c) and bottom (Au, d) layers. (e) Top- and bottom-layer absorption evaluated at different wavelengths (400 nm, 630 nm) for the hetero-metallic bi-layer NPM and its homo-metallic counterpart.

The experimental investigation of the plasmonic properties of the bi-layer porous film was performed using a number of techniques. First, the optical properties of the film were studied by means of spectroscopic ellipsometry. Fig. 3 illustrates the results for the Au-Ag system, while data for the individual layers are reported in the SI. The ellipsometrical analysis reveals spectroscopic features associated to an effective dielectric permittivity that significantly differs from that of the constituent materials (Fig. 3a). Such a quantity has been calculated for the nanoporous films through the fitting of the ellipsometrical angles Ψ and Δ (Fig. 3b,c) using a convolution of oscillators whose parameters are listed in Table 1.

**Table 1**: Parameters used to carry out the fit of the ellipsometrical angles.

| Oscillator Type | $A_i$ | $E_0$ (eV) | $\gamma_i$ (eV) | $E_g$ (eV) |
|---|---|---|---|---|
| Gaussian | 2.13 | 1.86 | 0.42 | n.a. |
| Gaussian | 1.08 | 2.57 | 1.32 | n.a. |
| Tauc-Lorentz | 0.3 | 3.5 | 0.45 | 1.87 |
| Gaussian | 1.06 | 4.3 | 0.41 | n.a. |

The Gaussian oscillators are characterized by an imaginary part in the form given in Eq. 1:

$$Im\{\varepsilon_{Gauss}\} = A_i e^{\left(-\frac{E-E_0}{\gamma_i}\right)} \quad (1)$$

where $E_0$ is the central energy of the $i_{th}$ oscillator, γ its damping and A its amplitude. The imaginary part of the Tauc-Lorentz oscillators is, instead, provided in Eq. 2:[25]

$$Im\{\varepsilon_{Tauc-Lor.}\} = \frac{A_i E_0 \gamma_i (E-E_g)^2}{(E^2-E_0^2)^2 + \gamma_i^2 E^2} \frac{1}{E}, E > E_g; \; Im\{\varepsilon_{Tauc-Lor.}\} = 0, E \leq E_g; (2)$$

Here, the parameter $E_g$ represents the optical band-gap. The associated real parts have been retrieved via Kramers-Kronig relations.

The Ag-Au nanoporous bi-layer manifests an Epsilon-Near-Zero (ENZ) transition at ~ 387 nm (Fig. 3a), significantly shifted from the natural ENZ wavelength occurring for pure Ag at the so-called Ferrell-Berreman mode around 327 nm.[26]–[28] Noticeably, the imaginary part of the effective dielectric permittivity at the ENZ transition of the multilayer is about 0.6, confirming its low-loss nature of this transition. Far from 387 nm, the nano-porous multilayer showcases the typical features of a noble metal with a large negative real dielectric permittivity associated to an increasing imaginary part. In addition, an absorbance peak is found at 663 nm, well captured by the Tauc-Lorentz oscillator that reveals an optical bandgap at 1.87 eV (663 nm, see Table 1). These results confirm the capability of the multilayered nano-porous architecture to enable the design of ENZ transitions and optical bandgaps far from those related to the constituent materials. It is also worth noticing that the Ag-Au system displays high electron mobility, as proved by pump-probe spectroscopy measurements (see SI-note#3). The fast decay time due to electron-electron and electron-phonon interactions is of the order of 0.5 ps, thus confirming that this system can display the same very high conductivity even when a hybrid Ag-Au multilayer is created, for instance compared to pure Au nanoporous films.[29]

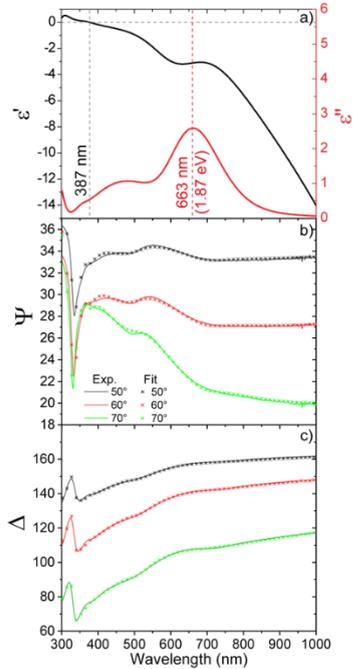

**Fig. 3:** (a) Real (black curve) and imaginary (red curve, associated right axis) part of the effective dielectric permittivity of the Ag-Au nanoporous bi-layer. Measured (solid lines) and fitted (asterisks) ellipsometrical angles Ψ (b) and Δ (c).

Spectroscopic ellipsometry demonstrated how a bi-layer structure can modify the optical properties of the NPM film. As already mentioned, NPMs, connected metallic nanomaterials, are mainly exploited for the strong local electromagnetic (EM) fields over a broad optical range. These fields are confined to subwavelength volumes and are termed hotspots. Due to diffraction-limited spatial resolution, however, far-field optical measurements cannot detect these hotspots. On the contrary, near-field techniques, like scanning near-field optical microscopy (SNOM), can be used to map the EM fields with high spatial resolution. Several reports indicate that SNOM analyses of NPMs can be a powerful tool to see the localized fields within the ligaments.[30], [31] The main limitation in SNOM experiments is related to the discrete energy of optical excitation that can be used (in our experiment we performed SNOM maps over the NPM film exciting at 550 nm – see Supporting Information – Note #2). On the contrary, Electron microscopy techniques such as electron energy loss spectroscopy (EELS) and cathodoluminescence (CL), enable to study the excitation from an interaction with a highly localized (subnanometer) electron beam that can be considered as a "white source" comprising a broad range of frequencies and is therefore able to excite all available modes in the sample. In a raster-scanning geometry, it is possible to retrieve the full spectrum from each pixel of the entire scanned sample with nanometer resolution.[32], [33] EELS measures the energy transfer from electrons to plasmons by calculating the energy loss of electrons that have passed through the sample. CL, on the other hand, directly collects photons scattered by the sample, reflecting the efficiency of the decay of plasmons into radiation. Thus, in both techniques, it is possible to map the local electric fields with nanoscale resolution. EELS and CL investigations on NPMs were reported, mainly focussing on nanoporous Au and nanoporous Ag as single layer films.[34] Here, we used both EELS and CL to map the local electric fields of the bi-layer nanoporous film and we compared the obtained results in order to better understand the behaviour of the bi-metallic system.

EELS is a technique that can probe the local optoelectronic properties of a material by using an electron beam to measure the energy each electron loses by interacting with the sample. This technique operates in transmission, and it works best on very thin samples, where the electrons ideally interact only once with the target material.

For EELS measurements, the nanoporous Ag-Au bi-layer film was directly fabricated onto a silicon nitride membrane (30 nm thickness), ensuring a low contribution to the background. The nanoporous nature of the sample leads to non-uniform coverage of the supporting film, as seen in Figure 4, including some gaps (bright areas). The study of the EELS signal was restricted to these areas by applying a mask over the spectrum image, in a similar approach to routine analysis of plasmons in isolated nanoparticles. The relevant regions were identified by applying a threshold to the zero-loss peak (ZLP) intensity. The ZLP contains all the electrons that have interacted elastically or quasi-elastically, and is highly sensitive to the total thickness the electrons travel through. The ZLP tail was then removed using a power law fit. The resulting hyperspectral image contains extensive information, including contributions from several energy loss mechanisms.

To unravel the signal, we adopted a multivariate analysis approach through Non-negative Matrix Factorisation (NMF).[35] This decomposition produces a set of components, without specific shape constraints, by analysing local correlations of the spectra, and is suitable for spectral features that can present a structure with high-order associated peaks. Furthermore, the components are built following statistical criteria and with minimal operator input, preventing human bias. In this case the NMF algorithm returns four components, displayed in Figure 4b-c. As usually observed for plasmon excitations using EELS, they typically present a peak at low energies followed by a secondary peak at roughly double the energy. Most of the features are below the visible range, accessing a region of the spectrum that is complementary to the CL observations. To map the local excitations, color-coded maps were constructed from the NMF components (Figure 4b). The colour for a given pixel represents the weight of each component. As can be expected from the complex geometry of the film, the components have a complicated spatial distribution with significant overlap, clustering around some local features. For example, some components are particularly dominant along edges or in the centre of cavities. To show the local energy loss phenomena on an area with a defined geometry, a Y-shaped cavity is chosen and presented in detail in Figure 4e-i. For ease of visualization the total intensities within a spectral (energy) window are shown.

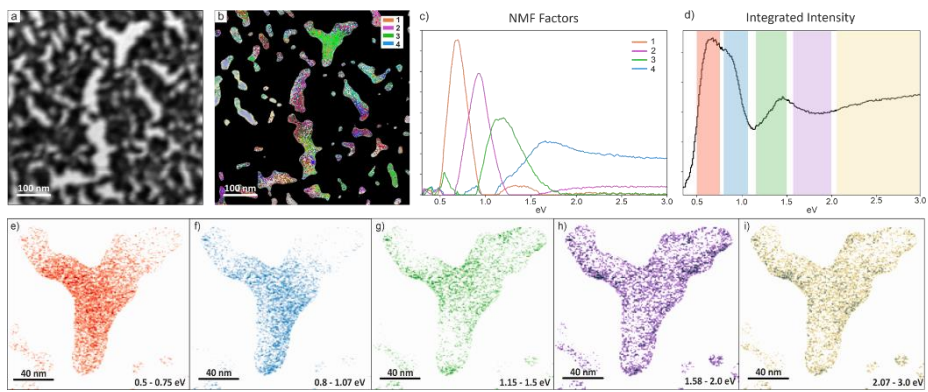

**Fig. 4:** EELS measurements over the nanoporous film. (a) Reference ADF (Annular Dark Field) image – note that contrast is reversed due to the total thickness of the film. (b) Spatial distribution of the NMF components and (c) their profiles. (d) Integrated EELS signal and (e) integrated intensities of the EELS signal within various energy windows, chosen in correspondence of the strongest signal associated with each component.

The windows were chosen by selecting the energy range in which each component was strongest, centering around the maximum intensity and taking a width corresponding to the full width at half maximum of the dominant peak. The resulting maps clearly show the presence of different harmonics that can be attributed to collective plasmon excitations of the complex geometry, including one concentrated in the center, one with hot spots along the edges and other, higher order excitations. In order to visualize plasmon contributions within the visible range, a window was selected in the range from 2.05 to 3.0 eV, roughly corresponding to the wavelength range studied with CL. At this energy, the EELS signal distribution replicates well the one at half the energy, corresponding to the highest order plasmons, concentrated mostly at the edges of Au/Ag network. It should be noted that the NMF decomposition was carried out again, restricting the signal within this area, in order to highlight potential excitation modes that were highly localized and could have been missed by the overall NMF over the entire hyperspectral image; the resulting components were qualitatively identical.

Figure 5 depicts the CL properties of the Au-Ag nanoporous films presented in Figure 5a. The panchromatic optical response, which represents the total light intensity captured during the electron-beam's dwell time, is mapped onto each pixel. CL emission over subwavelength distances is due to excitation of surface plasmon (SP) modes of the metal network. The Au-Ag nanoporous film produces a multitude of localized emission spots, showcasing its potential for various functionalities. To gain a deeper understanding of the Au-Ag nanoporous network's optical properties, a few spectra were extracted from specific locations on the network, as indicated by the colored crosses in Figure 5a. The spectra revealed the presence of SP modes exhibiting various intensities and energies across the visible (VIS) spectrum (Figure 5c). The emission at wavelengths shorter than 500 nm, which is below the energy threshold for interband transitions in gold, can be attributed to the presence of silver in the nanoporous film. As also can be noted in Figure 5a,b, SP modes appear mainly from metal clusters; however, faint optical emission comes from empty regions, probably due to plasmon coupling through the metal network.[32]

To visualize the distribution of surface plasmon modes across the nanoporous film, we created a color-coded map (Figure 5d) that assigns the specific wavelength of maximum CL to each pixel in the panchromatic map of Figure 5b. This map reveals the variation in SP mode energies across neighboring pixels, reflecting the short-range variation of the Au-Ag network due to the interplay of metal clusters and empty regions. It is worth noting that pixels with weak CL peaks below a few thousand counts are displayed in dark green and brown colors, these pixels correspond to the dark grey and black regions reported in Figure 5b.

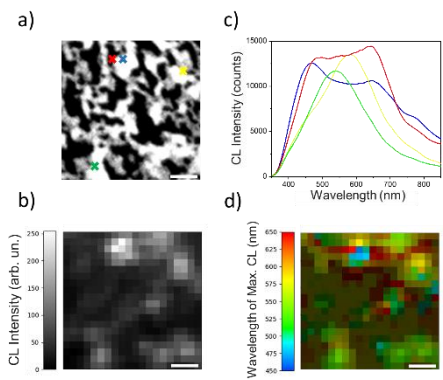

**Fig. 5:** (a) Scanning electron microscope (SEM) image (500 nm × 500 nm) of the studied nanoporous network, acquired simultaneously with CL acquisition. (b) Panchromatic CL map of the network in a) over the wavelength range of 350-850 nm. (c) Point CL spectra extracted from a few probe positions on the Au-Ag network, indicated by the corresponding colored crosses in a). (d) Color-coded map showing the wavelength of maximum CL for each pixel of the panchromatic map in b). Pixels with weak CL peaks below a few thousand counts are represented in dark green and brown colors. These pixels correspond to the dark gray and black areas in b). Scale bars are 100 nm long.

## 3 Conclusions

The results shown in this paper demonstrate how it is possible to easily prepare multilayers of NPMs using a dry process. The method can be extended to generic combinations of metals that can be endowed with a nanoporous morphology. The proof-of-concept reported here for a bi-layer of Ag on Au demonstrated how the plasmonic properties of the obtained nanoporous films can lead to significantly different effective dielectric permittivities than those of the original constituent metals, allowing to produce nanoporous metallic layers with tailored optical response. The possibility to both access and tune special features like ENZ dispersion is demonstrated through spectroscopic ellipsometry analysis. The pivotal role exerted by the localized electric field through the excitation of hot spots is highlighted both by EELS experiments, SNOM measurements and numerical calculations while CL observations allowed to spatially and spectrally map the radiative relaxation dynamics of the excited plasmons. The technique showcased in the manuscript can also be extended to more sophisticated configuration made of many different metallic layers stacked on top of each other.

## 4 Materials and Methods

*Samples preparation:*
The samples preparation is based on the original methods proposed in ref.[10]. In brief, Poly(methyl methacrylate) (PMMA) was spin-coated on a suitable substrate (Silicon for SEM and elipsometry, thin SiN membrane for EELS and CL, fused silica for ultrafast dynamics) at 4000 rpm for 2 min. Each metal (>99.99% purity) was evaporated by e-beam on a PMMA thin film at room temperature with an oblique angle of 80°, a rate of 0.1 nm/s, a target thickness of 12 nm was used in all the cases. The deposited Au film was plasma treated in $O_2$ with 200 W till the whole PMMA layer was removed. The Ag film was plasma treated in N2 with 200 W till the whole PMMA layer was removed. The bi-layer was obtained repeating the preparation using the Au porous layer as substrate for the preparation of the Ag porous layer on top.

*Spectroscopic Elipsometry:*
Spectroscopic ellipsometry was carried out by a M2000 apparatus by Woollam. Spectroscopic analysis was carried out at five different angles (30°, 40°, 50° and 60°). The fitting procedure was carried out by starting with a point-by-point fit of the pseudo-dielectric permittivity obtained directly from the measured ellipsometrical angles. The obtained imaginary dielectric permittivity was then fitted as a convolution of oscillators, as described in the main manuscript, to obtain a Kramers-Kronig-consistent analytic expression for the effective dielectric permittivity of the NPMs. The real parts of the dielectric permittivities have been obtained by applying Kramers-Kronig relations to the imaginary parts.

*Numerical simulations:*
To explore the plasmonic properties of porous NPMs, numerical investigation of the electromagnetic response of such material can be conducted using a finite element method (FEM) commercial software. COMSOL Multiphysics has been employed to model the optical behavior of the nanoporous metal. In particular, following a procedure reported in details in our recently reported works[7], [23], nanometric pores and irregularities of NPMs have been numerically reproduced by means of SEM images from experimental samples.

*SNOM measurements:*
SNOM measurements have been carried out on a side angle configuration. A 532 CW laser was focused directly into a hollow tip (60 nm diameter) through a 20x objective on a WITec confocal microscope. The near field signal was images through a side angle camera

*EELS measurements:*
The films were directly deposited on silicon nitride chips with a 30 nm membrane. Analysis was carried out at 60kV acceleration voltage in a ThermoFisher Spectra300 S/TEM equipped with a monochromator with "UltiMono". The energy resolution of the spectra was ~80 meV. The data were analysed using Hyperspy, a python-based data analysis toolkit.

*Cathodoluminescence measurements:*
Nanoporous Au-Ag films were fabricated on a 100 nm thick $Si_3N_4$ membrane to significantly minimize emission from bulky substrates during cathodoluminescence (CL) investigation. CL analysis was performed at room temperature by using a Zeiss Merlin scanning electron microscope (SEM) equipped with a high-performance CL imaging system (SPARC from Delmic). CL was spectrally resolved in the range of 350-850 nm with an "Andor Kimera 193i" spectrometer with a focal length of 193 mm and a grating of 150 gr/mm. The photon emission was captured by an "Andor Zyla5.5" CMOS camera with a maximum quantum efficiency of 60%. The electron beam operated at an acceleration voltage of 30 kV and an emission current of 10 nA. The panchromatic CL map consisted of 20 × 20 pixels (pixel size: 25 nm). The focused electron beam was scanned across the specimen, dwelling for 10 s (integration time) on each pixel to acquire the CL spectra. The intensity in each pixel of the panchromatic map corresponds to the integrated intensity of the detected light during the integration time.

*Pump-probe experiments*
Transient transmission measurements are carried out with a home-built spectroscopy system based on a commercial Yb:KGW regenerative amplifier system at a laser repetition rate of 50 kHz. A noncollinear optical parametric amplifier (NOPA) working in the VIS spectral range generates the probe pulse (time duration 12 fs). Off-resonant pump beam is generated from another NOPA working in the VIS/NIR spectral range centered at 800 nm (pulse duration 9 fs). Pump and probe pulses are temporally compressed by custom-designed dielectric chirped mirrors and are focused onto the sample noncollinearly, in order to spatially block the pump pulse after sample interaction. Residual scattered pump radiation is further spectrally suppressed with a dielectric long-pass filter (Thorlabs, FELH600). Spectrally resolved detection of the probe pulse after sample interaction is achieved by using a spectrograph (Acton) in combination with a high-speed charge coupled device camera operating at 50 kHz. Finally, a Pockels cell modulates the pump pulse train at half the repetition rate of the laser system, allowing the calculation of ΔT/T on a 25 kHz basis.

**Research funding:** The authors thank the European Union under the Horizon 2020 Program, FET-Open: DNA-FAIRYLIGHTS, Grant Agreement 964995 and the HORIZON-Pathfinder-Open: 3D-BRICKS, grant agreement 101099125. V. C. thanks the research project "Componenti Optoelettronici Biodegradabili ed Eco-Sostenibili verso la nano-fotonica "green"" (D.M. n. 1062, 10.08.2021, PON "Ricerca e Innovazione" 2014-2020), contract identification code 1062_R17_GREEN. TT, HL and NM acknowledge support from the Swedish Research Council (grant no. 2021-05784), Kempestiftelserna (grant no. JCK-3122) and the European Innovation Council (grant no. 101046920 'iSenseDNA').

**Author contribution:** VC performed the optical spectroscopy, HY and PF developed the dry-method, TT, HL and NM performed the ultra-fast dynamics characterization, AS and AA performed the numerical simulations, MC, GB performed the CL characterization, ADL performed the SNOM analysis, RK supported the activity, GD, YI and AG performed the EELS characterization, DG conceived the idea and supervised the work

All authors have accepted responsibility for the entire content of this manuscript and approved its submission.

**Conflict of interest**: Authors state no conflict of interest.